\documentclass{jfmc}

\usepackage{graphicx}
\usepackage{newtxtext}
\usepackage{newtxmath}
\usepackage{natbib}
\usepackage{hyperref}
\usepackage[normalem]{ulem}
\hypersetup{
    colorlinks = true,
    urlcolor   = blue,
    citecolor  = black,
}

\newcommand{\RomanNumeralCaps}[1]
\linenumbers

\title{Self-similar fault slip in response to fluid injection}

\author{Robert C. Viesca\aff{1}
  \corresp{\email{robert.viesca@tufts.edu}}}

\affiliation{\aff{1}Department of Civil and Environmental Engineering, Tufts University, Medford, MA 02155, USA}

\begin{document}
\maketitle

\begin{abstract}
There is scientific and industrial interest in understanding how geologic faults respond to transient sources of fluid. Natural and artificial sources can elevate pore fluid pressure on the fault frictional interface, which may induce slip. We consider a simple boundary value problem to provide an elementary model of the physical process and to provide a benchmark for numerical solution procedures. We examine the slip of a fault that is an interface of two elastic half-spaces. Injection is modeled as a line source at constant pressure and fluid pressure is assumed to diffuse along the interface. The resulting problem is an integro-differential equation governing fault slip, which has a single dimensionless parameter. The expansion of slip is self-similar and the rupture front propagates at a factor $\lambda$ of the diffusive lengthscale $\sqrt{\alpha t}$. We identify two asymptotic regimes corresponding to $\lambda$ being small or large and perform a perturbation expansion in each limit. For large $\lambda$, in the regime of a so-called critically stressed fault, a boundary layer emerges on the diffusive lengthscale, which lags far behind the rupture front. We demonstrate higher-order matched asymptotics for the integro-differential equation, and in doing so, we derive a multipole expansion to capture successive orders of influence on the outer problem for fault slip for a driving force that is small relative to the crack dimensions. Asymptotic expansions are compared to accurate numerical solutions to the full problem, which are tabulated to high precision.
\end{abstract}

\section{Introduction}
\label{sec:intro}

\medskip

The coupling of fluid flow and fracture of an elastic medium has an extensive history in the context of open-mode fractures, specifically hydraulic fracture, in which an injected fluid drives crack opening. Early work presumed laminar flow of the fluid in a planar crack \citep{Khristianovic55, Barenblatt56} leading to similarity solutions and crack-tip asymptotics \citep{Spence85, Desroches94} that depart from the square-root asymptotic behavior of classical linear elastic fracture mechanics. Further developments include explicit consideration of fluid leak-off into the elastic medium \citep{Lenoach95,Adachi08}, the possibility of fluid lagging behind the crack tip \citep{Garagash00}, and turbulent flow \citep{Lister90b,Tsai10}, with \citet{Detournay16} providing a more complete review of related progress. In addition to applications to oil and gas extraction from subsurface reservoirs, such models have also been widely applied to magmatic intrusions \citep{Lister90a, Lister90b, Rubin95, Bunger11, Michaut11}.

\medskip

The interplay between fluid flow and elasticity has received renewed interest in a comparable problem of fluid-driven delamination of a thin elastic sheet from a rigid substrate. In this problem, the sheet's elastic response is represented by classical beam theory, such that elastic interactions are local and solutions are found in a relatively straightforward manner. In comparison, hydraulic fracture in a full space, in which elastic interactions are non-local, reduces to singular integro-differential equations, requiring more specialized solution techniques. A peculiarity of the thin-sheet problem is that the inherent neglect of variations over distances on the order of the sheet thickness necessitates regularization of fluid flow near the fracture tip \citep{Flitton04}. This regularization may either take the form of a fluid lag behind the tip \citep{Hewitt15,Ball18, Wang18} or a pre-existing thin-film of fluid, such that a rupture front is no longer precisely defined \citep{Flitton04,Hosoi04,Lister13,Hewitt15,Peng20}, or may be bypassed altogether by explicit consideration of near-tip phenomena over distances comparable to the sheet thickness \citep{Lister19}.

\medskip

We now look to examine the shear-fracture counterpart to the above problems, in the context of fluid-induced slip of geologic faults, the thin-sheet limit of which is of interest in the modeling of landslides \citep{Palmer73,Puzrin05,Viesca12}, snow-slab avalanches \citep{McClung79}, and short-timescale ice-sheet motion \citep{Lipovsky17}, among other problems. A source of fluid in the subsurface can drive the sliding-mode fracture of a geologic fault by locally reducing the fault's frictional shear strength below an ambient level of shear stress such that fault must slide. The fluid source may be natural, such as from mineral dehydration of subducted sediments \citep{Peacock01} or from a mantle source \citep{Kennedy97}, or artificial, such as from the subsurface injection of fluids at kilometer scale depths for the disposal of wastewater \citep{Healy1968} or for the enhancement of permeability in geothermal, oil, or gas reservoirs. Despite wide interest, simple solutions for fluid-induced fault slip are scarce.

\medskip

In the context of frictional fracture, much of the development of fluid-fracture interaction has focused on geologic fault phenomena, including earthquakes and slow, aseismic slip. Fluid-driven fault slip has been studied in the context of earthquake nucleation via the initiation of aseismic slip \citep{Viesca12,Garagash12,Jha14,Ciardo19,Zhu20,Garagash21}, or strictly stable, aseismic slip \citep{Rutqvist07,Garipov16,Bhattacharya19,Dublanchet19,Yang21}. In addition, there has been substantial focus on fluid-involved feedback mechanisms during seismic or aseismic fault rupture, in which the role of fluids is in response to sliding such that rupture is fluid-assisted or fluid-inhibited. These including dilatancy \citep{Rice73,Yang21,Brantut21}, thermal pressurization by frictional heating \citep{Rice06,Noda09,Schmitt11,Garagash12a,Viesca15} and chemical decomposition \citep{Platt15}. Within this body of work, solutions are nearly entirely numerical.

\medskip

We examine a model for fluid-driven fault rupture, in which the condition of fluid flow is rudimentary but physically plausible and comparable to the starting assumption of laminar flow in hydraulic fracture. We consider a planar fault in an unbounded, linear-elastic medium under a uniform state of stress prior to injection. Quasi-static deformation is in-plane or anti-plane, such that the corresponding fault slip is mode II or mode III rupture. We use a boundary integral formulation to relate the crack-face displacement and traction. Injection is modeled as a line source of fluids at constant pressure following which fluid migration, and the concomitant rise in pore fluid pressure, is restricted to occur along the fault plane. The fault strength is frictional and is the product of a constant coefficient of friction and the local effective normal stress, the difference between the fault-normal traction and the local pore fluid pressure. 

\medskip

This problem was presented in \citet{Bhattacharya19} and is a variation of one considered in depth by \citet{Garagash12} (hereafter referred to as GG12). GG12 examined the response to injection of a fault under the same elastic and fluid conditions considered here. However, GG12 considered a fault friction coefficient that weakens with slip, a feature that gives rise to rich behavior, including the possibility of dynamic rupture nucleation and arrest, corresponding to an earthquake source. GG12 found two end-member regimes corresponding to marginally pressurized and critically stressed faults, which reflect the pre-injection state of stress. The authors showed that these regimes lead to a rupture front lagging or outpacing fluid diffusion, respectively, and that, in the critically stressed regime, fault slip can be described by a boundary-layer analysis. However the slip-dependent strength in that problem required numerical solution, even in end-member regimes, and growth of the slipping region was non-trivial and dependent on at least two problem parameters. 

\medskip

A constant Coulomb friction coefficient necessitates stable growth of fault rupture. Furthermore, the propagation of fault slip occurs in self-similar manner in response to a self-similar source. Additionally, the problem has a single dimensionless parameter, with the same two end-member regimes as in the problem considered by GG12. The simpler problem admits closed-form perturbation expansions in these regimes, and readily allows for higher-order asymptotic matching of the boundary-layer problem in the critically stressed regime. In this regime, where the rupture front races ahead of the elevated fluid pressure distribution, the problem provides for the development of a multipole expansion to consider the higher-order source effects. The problem is also amenable to accurate numerical solutions for the cases intervening the two end-member regimes and we provide tabulated solutions to high precision.

\medskip

We begin in Sections \ref{sec:2} and \ref{sec:3} by providing a problem statement and summary of asymptotic solutions to leading order in the critically stressed and marginally pressurized limits. Subsequently, we return to the full problem and summarize its solution in Section \ref{sec:4}. Finally, we revisit the end-member regimes and derive the asymptotic solutions to second order. In the marginally pressurized limit, the solution can be written as a single perturbation expansion (Section \ref{sec:5}). In the critically stressed limit, inner and outer perturbation expansions are found and the two solutions are matched to construct a composite solution (Sections \ref{sec:6}--\ref{sec:8}). We compare the asymptotic expansions to numerical solutions to the full problem throughout.

\begin{figure}
\centerline{\includegraphics[scale=0.75]{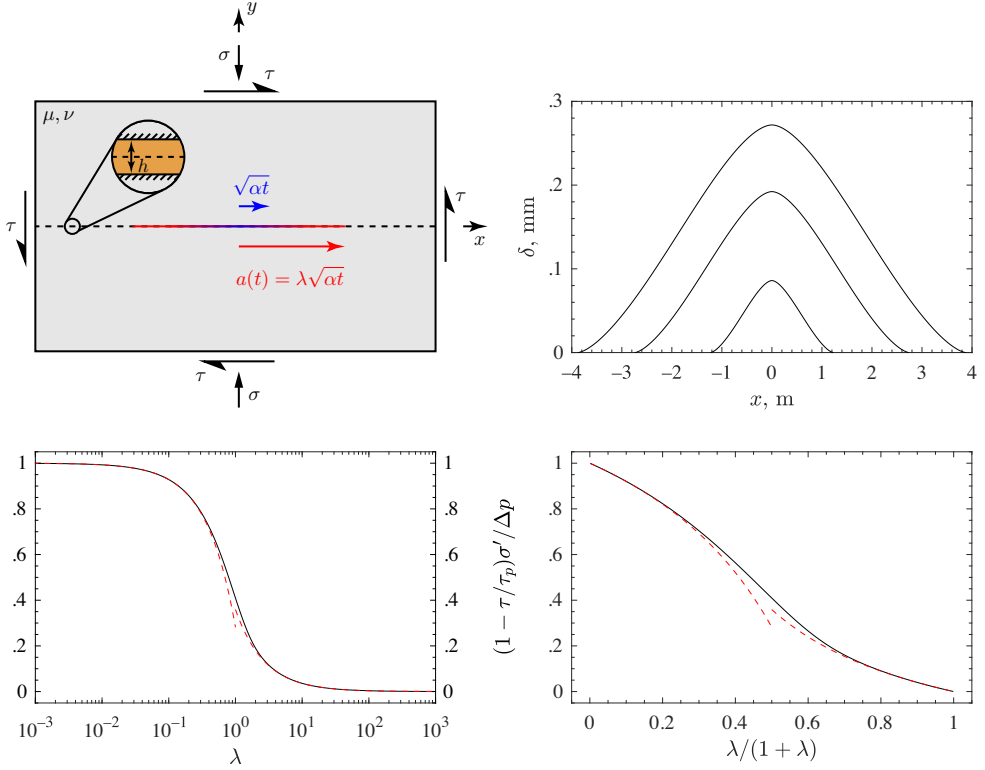}}
\caption{{\bf Counter-clockwise:} ({\bf top left}) Unbounded elastic body containing a fault, loaded remotely with fault-normal and shear stress $\sigma$, $\tau$. The fault is embedded within a thin poroelastic layer, assumed to be much more permeable than surroundings. Fluid injected at $x=0$ diffuses along fault as $\sqrt{\alpha t}$, inducing quasi-static slip out to a distance $a(t)$. Fault has constant friction coefficient $f$. ({\bf bottom left}) Black: relation between rupture growth factor $\lambda$ and a parameter reflecting the initial state of stress and injection pressure, where $\sigma'=\sigma-p_o$ and $p_o$ is pre-injection fault fluid pressure. Dashed: asymptotic behaviors, eqs. (\ref{eq:asy1}) and (\ref{eq:asy2}). ({\bf bottom right}) Same as bottom left, with abscissa arranged to occupy a finite interval. ({\bf top right}) Plot of self-similar slip distributions at three instants in time after the start of injection, $t=1$, 5, and 10 min., for the specific choices $\sigma=50\text{ MPa}$, $\tau=12\text{ MPa}$, $p_o=20\text{ MPa}$, $\Delta p = 12 \text{ MPa}$, $f=0.5$, $\alpha_{hy}=0.01 \text{ m}^2/\text{s}$, $\mu=30\text{ GPa}$, $\nu=1/4$, $\mu'=20\text{ GPa}$. For these choices, the parameter $(1-\tau/\tau_p)\sigma'/\Delta p=0.5$. The corresponding self-similar slip distribution and factor $\lambda$ are given in Table S1 in the supplementary materials.}
\label{fig1}
\end{figure}

\section {Problem formulation}
\label{sec:2}

\medskip

\subsection {Fluid mechanics}

\medskip

A planar fault slip surface along $y=0$ is modeled as an interface within a poroelastic layer of thickness $h$ that is itself embedded within an elastic body. The layer corresponds to a fault core that is presumed to be much more permeable than the surrounding host rock, but having comparable elastic properties. We assume that fluid flow within the layer follows Darcy's law and we examine the injection of fluid directly into the fault core as a source of constant pressure distributed across the layer thickness at $x=0$. For this poroelastic configuration, which is an in-plane version of an axisymmetric case considered by \citet{Marck15}, the pore fluid pressure distribution is uniform across the layer thickness and its distribution $p(x,t)$ along the fault coordinate $x$ satisfies a diffusion equation
\[ p_{t} =\alpha_{hy} \, p_{xx} \]
where $\alpha_{hy}$ is the hydraulic diffusivity of the fault core and where the pore pressure is subject to the conditions of the initial state and injection at constant pressure $\Delta p$ at $x=0$, 
\[ \quad p(x,0)=p_o,\quad p(0,t>0)=\Delta p \]
the known solution to which is
\begin{equation}p(x,t)=p_o+\Delta p\, \text{erfc}(|x|/\sqrt{\alpha t})\label{eq:dp}\end{equation}
where we adopt a nominal diffusivity 
\[ \alpha=4\alpha_{hy} \]
in which the hydraulic diffusivity $\alpha_{hy}=k/(\beta \eta)$, where $k$ is the Darcy permeability of the layer, $\eta$ is the viscosity of the permeating fluid, and $\beta$ is a storage coefficient reflecting the compressibility of the fluid and porous matrix.
As noted by \citet{Marck15}, when the diffusive distance $\sqrt{\alpha t}$ is much larger than the layer thickness $h$, the local response of the poroelastic layer is effectively that of uniaxial vertical strain in proportion to the local pore fluid pressure.

\subsection{Solid mechanics}

\medskip

We assume that the poroelastic layer thickness $h$ is small such that the condition $\sqrt{\alpha t}\gg h$ is quickly achieved. In the case, the presence of the layer, apart from its role in conducting fluids, need not be explicitly considered and the shear and normal traction conditions on the layer-medium boundary may be directly applied to the sliding interface. Furthermore, provided $\sqrt{\alpha t}$ is sufficiently greater than $h$, we may reasonably neglect any fault-normal stress changes due to the swelling of the poroelastic layer about the source  \citep{Marck15}. The medium containing the fault is assumed to be linearly elastic and its deformation, along with slip on the fault, may be in-plane or anti-plane. The in-plane case is illustrated in Fig. \ref{fig1} (top left). The shear modulus of the medium is $\mu$ and the Poisson ratio $\nu$. We define the effective elastic modulus $\mu'=\mu/[2(1-\nu)]$ for in-plane (mode-II) case and $\mu'=\mu/2$ for anti-plane (mode-III) case. We denote the initial (pre-injection) fault shear stress $\tau$ (in-plane or anti-plane), the fault friction coefficient $f$, the initial total fault-normal compressive stress $\sigma$, and the initial effective normal stress $\sigma'=\sigma-p_o$, where $p_o$ is the pre-injection pore fluid pressure in the layer. The initial fault strength is $\tau_p=f\sigma'$.

\medskip

The fault obeys a Coulomb friction law: the local shear strength of the fault $\tau_s$ is a constant proportion of the local effective normal stress, with a constant coefficient of friction $f$
\begin{equation}\tau_s(x,t)=f[\sigma-p(x,t)]\end{equation}
Where sliding occurs, this strength must equal the shear stress on the fault. The shear stress can be decomposed into a sum of the initial shear stress $\tau$ plus quasi-static changes due to a distribution of slip $\delta$ \citep[e.g., ][]{Rice68}, such that the stress-strength condition is
\begin{equation}\tau_s(x,t)=\tau+\frac{\mu'}{\pi}\int_{-a(t)}^{a(t)} \frac{\partial \delta(s,t)/\partial s}{s-x}ds\label{eq:tau}\end{equation}
where $x=\pm a(t)$ are the crack-tip locations.

\begin{figure}
\centerline{\includegraphics[scale=.75]{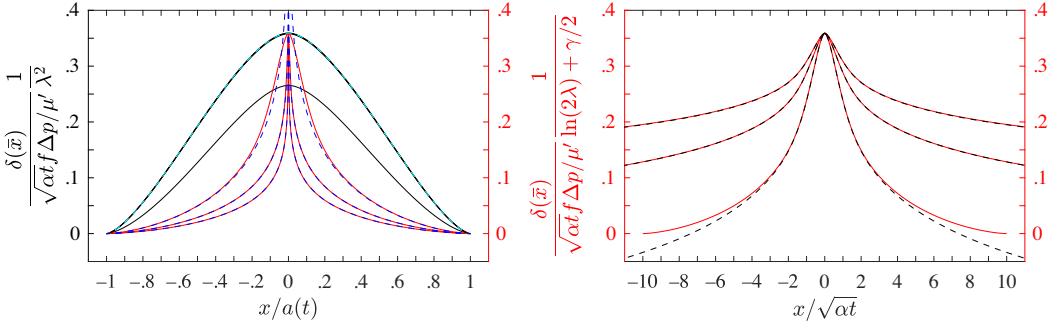}}
\caption{({\bf left}) Self-similar distributions of slip $\delta$ with distance from injection point $x$, which is scaled by the crack half-length $a(t)=\lambda\sqrt{\alpha t}$. Solid red and black curves are numerical solutions to the full problem. Dashed curves are leading-order asymptotic solutions. Each curve corresponds to one value of $\lambda$ in the range $\lambda=10^{-3},\,10^{-2},\,...\,,10^3$. Black curves correspond to $\lambda=10^{-3},\,10^{-2},\,10^{-1},\,10^0$ from top to bottom, with the first three indistinguishable on this scale; red curves correspond to $\lambda=10^1,\,10^2,\,10^3$ from top to bottom. Cyan-dashed: solution for small $\lambda$, eq. (\ref{eq:lsm}). Blue-dashed: ``outer" solutions for large $\lambda$, eq. (\ref{eq:out}). To facilitate comparisons, two vertical scales are used: one for black and cyan-dashed curves, and another for red and blue-dashed curves. ({\bf right}) For large values of $\lambda$, the distribution of slip is plotted over distances scaled by $\sqrt{\alpha t}$, which is much smaller than the crack length $a(t)$. This ``inner" behavior is described by eq. (\ref{eq:in}), a single numerical solution shown here as black-dashed curves. Curves correspond to $\lambda=10^1,\,10^2,\,10^3$ from bottom to top.}
\label{fig2}
\end{figure}

\section{Summary of results}
\label{sec:3}

\medskip

After non-dimensionalizing, the problem is found to have the sole parameter
\begin{equation}\left(1-\frac{\tau}{\tau_p}\right)\frac{\sigma'}{\Delta p}\label{eq:par}\end{equation}
that is bounded between 0 and 1. The upper bound denotes a marginally pressurized fault, where the fluid pressure increase is just sufficient to initiate sliding: $f[\sigma-(p_o+\Delta p)]=\tau$. The lower bound denotes a critically stressed fault, where the initial shear stress is equal to the initial shear strength: $\tau=\tau_p$. 

\medskip

The solution consists of a self-similar distribution of slip, in which the crack front grows as 
\[ a(t)=\lambda \sqrt{\alpha t} \]
and the slip distribution can be written as 
\[ \delta(x,t)\Rightarrow \delta(\bar x) \]
where the similarity coordinate is
\[ \bar x = x/a(t) \]
The factor $\lambda$, to be solved for, determines whether the crack lags ($\lambda<1$) or  outpaces ($\lambda>1$) the diffusion of pore pressure, which stretches as $\sqrt{\alpha t}$. $\lambda$ depends uniquely on the sole parameter (\ref{eq:par}), and that dependence is illustrated in Fig. \ref{fig1}b and tabulated at the top of  Table S1 in the supplementary materials. The self-similar profile of slip, as it depends on $|x|/a(t)$, is also presented in the bottom of Table S1 for several values of the parameter (\ref{eq:par}). Scaled plots of the self-similar profile for various values of $\lambda$ are shown in Fig. \ref{fig2}. In the limit that the parameter (\ref{eq:par}) approaches its end-member values, closed-form expressions for $\lambda$ and $\delta$ are available and provided below to leading order, with detailed derivations in the Sections that follow.

\subsection {Marginally pressurized faults, $\tau\rightarrow f(\sigma'-\Delta p)$}

\medskip

In this limit, the parameter $(1-\tau/\tau_p)\,\sigma'/\Delta p\rightarrow 1$, the factor $\lambda \ll 1$ (i.e., the rupture lags the diffusion of pore fluid pressure), and the relation between the two follows the asymptotic expansion
\begin{equation}(1-\tau/\tau_p)\,\sigma'/\Delta p\approx 1-\frac{4}{\pi^{3/2}}\lambda - O(\lambda^3) \label{eq:asy1}\end{equation}
The slip distribution in this limit is
\begin{equation}
\delta(\bar x)\approx\frac{\lambda^2 \sqrt{\alpha t} f \Delta p }{\mu'} \frac{2}{\pi^{3/2}}\left(\sqrt{1-\bar x^2}-\bar x^2\text{atanh}\,\sqrt{1-\bar x^2}\right)+O(\lambda^4)
\label{eq:lsm}
\end{equation}
and the accumulation of slip at the center is
\begin{equation}
\delta(0)\approx\frac{2}{\pi^{3/2}}
\frac{\lambda^2\sqrt{\alpha t}f\Delta p}{\mu'}+O(\lambda^4)
\label{eq:d0s}\end{equation}
The derivation of this solution is detailed in Section \ref{sec:5}.

\begin{figure}
\centerline{\includegraphics[scale=0.75]{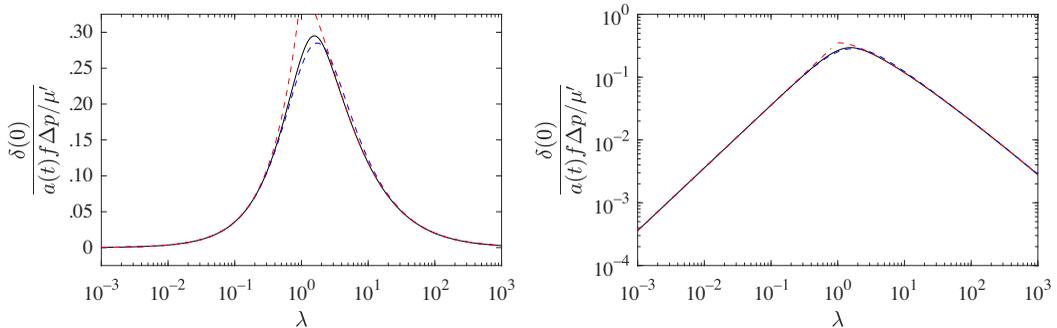}}
\caption{The maximum slip, which occurs at $x=0$, as it relates to the factor $\lambda$ shown as ({\bf left}) semi-log and ({\bf right}) log-log plots. Red-dashed: end-member scalings at small $\lambda$, eq. (\ref{eq:d0s}), and at large $\lambda$, eq. (\ref{eq:d0}). Blue-dashed: approximation of $\delta(0)$ for all $\lambda$, eq. (\ref{eq:d0approx}).}
\label{fig3}
\end{figure}

\subsection{Critically stressed faults, $\tau \rightarrow \tau_p$ }

\medskip

In this limit $(1-\tau/\tau_p)\,\sigma'/\Delta p\rightarrow 0$, $\lambda\gg 1$ (i.e., the rupture outpaces the diffusion of fluid pressure), and the asymptotic relation is
\begin{equation}(1-\tau/\tau_p)\,\sigma'/\Delta p\approx \frac{2}{\pi^{3/2}}\frac{1}{\lambda}+O(\lambda^{-3}) \label{eq:asy2}\end{equation}

\medskip

Similarly to the problem considered by \citet{Garagash12}, the solution for slip can be decomposed into an outer solution on distances comparable to the rupture distance $a(t)$, and an inner solution on distances comparable to the diffusion lengthscale $\sqrt{\alpha t}$. The two solutions are matched at an intermediate distance.

The outer solution for the slip distribution is
\begin{equation}
\delta (\bar x)\approx \frac{\sqrt{\alpha t} f \Delta p }{\mu'}  \frac{2}{\pi^{3/2}}\left(\text{atanh}\,\sqrt{1-\bar x^2}-\sqrt{1-\bar x^2}\right) +O(\lambda^{-2})
\label{eq:out}
\end{equation}
where $\bar x$ is the similarity coordinate used above. The derivation of this solution may be found in Section \ref{sec:6}.

\medskip

The inner solution is given by the expression
\begin{equation}
\delta(x/\sqrt{\alpha t}) \approx \delta(0) - \frac{\sqrt{\alpha t} f\Delta p}{\mu ' }\underline{ \int_0^{x/\sqrt{\alpha t}}\left[\frac{1}{\pi}\int_{-\infty}^\infty \frac{\text{erfc}(| \hat s |)}{\hat x-\hat s}d\hat s\right]d\hat x}+O(\lambda^{-2})
\label{eq:in}
\end{equation}
The underlined portion is evaluated numerically and provided as a supplementary function $f(x/\sqrt{\alpha t})$ in Table S2 with the similarity coordinate
\[\hat x = \frac{x}{\sqrt{\alpha t}} \]
For large distances $x/\sqrt{\alpha t}$, $f$ behaves as
\begin{equation}
f(\hat x)\approx\frac{2}{\pi^{3/2}}\left(\ln \left | \hat x \right | +\frac{\gamma}{2}+1\right)+O\left(\hat x^{-2}\right)
\label{eq:fbg}
\end{equation}
where $\gamma=0.57721566...$ is the Euler-Maraschoni constant. Using this asymptotic behavior to match the inner solution at large $x/\sqrt{\alpha t}$ with the outer solution at small $x/a(t)$ provides the slip at the center
\begin{equation}
\delta(0)\approx \frac{\sqrt{\alpha t}f\Delta p}{\mu'}\frac{2}{\pi^{3/2}}[\ln(2\lambda)+\gamma /2+O(\lambda^{-2})]
\label{eq:d0}
\end{equation}
in the large $\lambda$ limit. 

\medskip

Other properties of $f(x/\sqrt{\alpha t})$ include 
\[ f''(\hat x)=-\frac{2}{\pi^{3/2}}\exp(-\hat x^2) \text{Ei}(\hat x^2) \]
where $\text{Ei}(x)=-\int_{-x}^\infty \exp(-u)/u \,du$ is the exponential integral, and in the limit that $x/\sqrt{\alpha t}$ is small, $f$ behaves as
\begin{equation}
f(\hat x)\approx \frac{2}{\pi^{3/2}}\,\hat x^2\left(\ln \frac{1}{|\hat x|} -\frac{\gamma}{2} +\frac{3}{2} \right)+O\left(\hat x^4 \ln |\hat x|\right)
\label{eq:fsm}
\end{equation}
A detailed discussion of the inner solution and its matching to the outer solution, can be found in Sections \ref{sec:7} and \ref{sec:8}.

\subsection{Accumulation of slip at the injection point} 

\medskip

\begin{figure}
\centerline{\includegraphics[scale=.75]{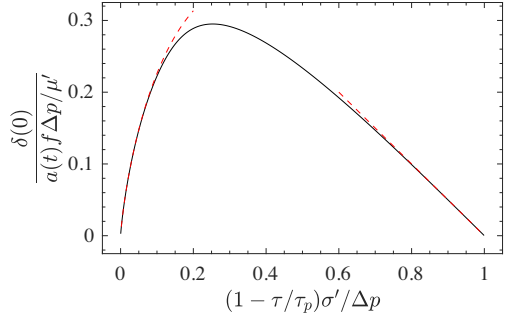}}
\caption{ The peak slip as it relates to the problem parameter $(1-\tau/\tau_p)\,\sigma'/\Delta p$. Red-dashed: scalings as the parameter approaches its bounds, derived from eqs. (\ref{eq:d0s}) and (\ref{eq:d0}) and the asymptotic relations between the parameter and $\lambda$, eqs. (\ref{eq:asy1}) and (\ref{eq:asy2}).}
\label{fig4}
\end{figure}

Figs. \ref{fig3} and \ref{fig4} show the solution for the peak slip, located at the injection point, as it depends on the parameter (\ref{eq:par}) or the factor $\lambda$. An approximation of peak slip at the injection point that respects the asymptotic behavior at both critically stressed and marginally pressurized limits---eqs. (\ref{eq:d0s}) and (\ref{eq:d0})---and is to within 5\% error over the intervening range of $\lambda$, is
\begin{equation}
\delta(0)\approx \frac{\lambda\sqrt{\alpha t} f \Delta p }{\mu'}  \frac{2}{\pi^{3/2}}\frac{\lambda}{1+\lambda^2/[\ln(6+2\lambda) +\gamma/2]}
\label{eq:d0approx}
\end{equation}

\section{Non-dimensionalization and solution to full problem}
\label{sec:4}

\medskip

Combining eqs. (\ref{eq:dp})--(\ref{eq:tau}) and rearranging leads to the non-dimensionalized equation
\begin{equation}
\left(1-\frac{\tau}{\tau_p}\right)\frac{\sigma'}{\Delta p}-\text{ erfc}\left| \lambda\bar x\right|=-\frac{1}{\pi}\int_{-1}^{1}\frac{d\bar\delta/d\bar s}{\bar x-\bar s}d\bar s
\label{eq:nond}
\end{equation}
where we have used $x=a(t)\bar x$,  $a(t)=\lambda\sqrt{\alpha t}$, and  $\delta(x,t)=\bar\delta[x/a(t)]a(t)f\Delta p/\mu'$. The solution we seek is the slip distribution $\bar\delta$ and the crack-growth prefactor $\lambda$, including their dependence on the problem parameter $(1-\tau/\tau_p)/\sigma'/\Delta p$. 

\medskip

We begin by looking for the solution for $\lambda$. To do so, we first note that to avoid a singularity in shear stress ahead of the crack tips, which is necessary because the Coulomb friction requirement implies a finite shear strength of the interface, the crack-tip stress intensity factors of the rupture must be zero. This condition implies (Appendix A) 
\begin{equation}
\left(1-\frac{\tau}{\tau_p}\right)\frac{\sigma'}{\Delta p}=\frac{1}{\pi}\int_{-1}^1\frac{\text{erfc}|\lambda x|}{\sqrt{1-x^2}}dx
\label{eq:Kint}
\end{equation}
which provides an implicit solution for $\lambda$ as it depends on the problem parameter. This relation is easily determined numerically since, for a given $\lambda$, the integrand on the right hand side can be evaluated by Gauss-Chebyshev quadrature (Appendix \ref{appB}). The behavior at large and small values of $\lambda$, eqs. (\ref{eq:asy1}) and (\ref{eq:asy2}), is found by asymptotic approximation of the integral.

\medskip

To solve for $\bar\delta$, we note that (\ref{eq:nond}) may be inverted for $d\bar\delta/d\bar x$ (Appendix A)
\begin{equation}
\frac{d\bar \delta}{d\bar x}=-\frac{\sqrt{1-\bar x^2}}{\pi}\int_{-1}^1\frac{\text{erfc}|\lambda \bar s|}{\sqrt{1-\bar s ^2}}\frac{1}{\bar x-\bar s}d\bar s
\label{eq:dinv}
\end{equation}
After having determined $\lambda$ via (\ref{eq:Kint}), the right hand side may be numerically evaluated and integrated to arrive to $\bar \delta(\bar x)$ using a Gauss-Chebyshev quadrature for singular integrals \citep{Erdogan73, Viesca18}. The numerical solution of (\ref{eq:Kint}) and (\ref{eq:dinv}) for $\lambda$ and $\bar \delta(\bar x)$ is detailed in Appendix \ref{appB}.
\begin{figure}
\begin{center}
\includegraphics[scale=.75]{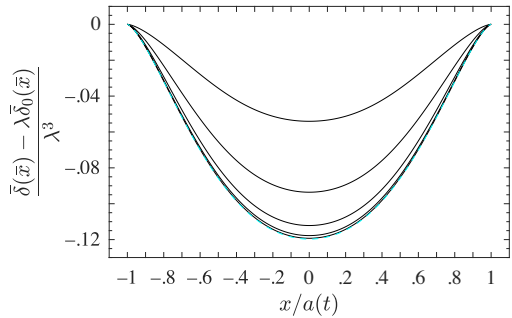}
\end{center}
\caption{Self-similar distribution of slip $\delta$, less the first-order term, eq. (\ref{eq:dMP1}), in the asymptotic expansion for slip eq. (\ref{eq:asyexp}) in the marginally pressurized limit (small $\lambda$). From top to bottom, black curves correspond to the difference for $\lambda=2,1,\frac{1}{2},\frac{1}{4},\frac{1}{8}$. Cyan-dashed curve is the second-order term of the expansion, eq. (\ref{eq:dMP2}) }
\label{fig5}
\end{figure}

\section{Solution in the marginally pressurized limit}
\label{sec:5}

\medskip

For $\lambda\ll 1$, we may use the expansion of the function
\begin{equation}\text{erfc}|\lambda \bar x|\approx 1-\frac{2}{\sqrt{\pi}}|\lambda \bar x|+\frac{2}{3\sqrt{\pi}}|\lambda \bar x|^3+...
\label{eq:exp1}
\end{equation}
to expand the integral in eq. (\ref{eq:Kint}) as
\begin{equation}
\left(1-\frac{\tau}{\tau_p}\right)\frac{\sigma'}{\Delta p}=\int_{-1}^{1}\frac{\text{erfc}|\lambda x|}{\sqrt{1-x^2}}dx\approx1-\frac{4}{\pi^{3/2}}\lambda +\frac{8}{9\pi^{3/2}}\lambda^3+O(\lambda^5)
\label{eq:exp2}
\end{equation}
In turn, we use eqns. (\ref{eq:exp1}) and (\ref{eq:exp2}) to reduce eq. (\ref{eq:nond}) to
\begin{equation}
\frac{1}{\pi}\int_{-1}^{1}\frac{d\bar\delta/d\bar s}{\bar x-\bar s}d\bar s \approx
\lambda\left(\frac{4}{\pi^{3/2}}-\frac{2}{\sqrt{\pi}}|\bar x|\right) + \lambda^3\left(-\frac{8}{9\pi^{3/2}}+\frac{2}{3\sqrt{\pi}}|x|^3\right)+O(\lambda^5)
\end{equation}
We write the solution to the above equation as the perturbation expansion
\begin{equation}
\bar \delta(\bar x)\approx \lambda\delta_0(\bar x)+\lambda^3\delta_1(\bar x)+O(\lambda^5)
\label{eq:asyexp}
\end{equation}
where $\delta_0$ and $\delta _1$ satisfy
\[ \frac{1}{\pi}\int_{-1}^{1}\frac{d\delta_0/d\bar s}{\bar x-\bar s}d\bar s=
\frac{4}{\pi^{3/2}}-\frac{2}{\sqrt{\pi}}|\bar x|,
\qquad \frac{1}{\pi}\int_{-1}^{1}\frac{d\delta_1/d\bar s}{\bar x-\bar s}d\bar s=
-\frac{8}{9\pi^{3/2}}-\frac{2}{3\sqrt{\pi}}|\bar x|^3 
\]
The solutions to which 
\begin{equation}\delta_0(\bar x)=\frac{2}{\pi^{3/2}}\left(\sqrt{1-\bar x^2}-\bar x^2 \text{atanh}\, \sqrt{1-\bar x^2}\right)
\label{eq:dMP1}\end{equation}
\begin{equation}\delta_1(\bar x)= \frac{1}{3\pi^{3/2}}\left((\bar x^2-2)\sqrt{1-\bar x^2}+\bar x^4\text{atanh}\,\sqrt{1-\bar x^2}\right)
\label{eq:dMP2}\end{equation}
are found by the linear superposition of particular solutions to the general problem
\[ g(x)=\frac{1}{\pi}\int_{-1}^1 \frac{h'(s)}{x-s}ds \]
that are provided in Table \ref{tab:kd} and found following Appendix A. Using (\ref{eq:asyexp}), the slip at the injection point in the marginally pressurized limit evaluates to 
\begin{equation}
\delta(0)\approx\frac{\lambda\sqrt{\alpha t}f\Delta p}{\mu'}\left(\frac{2}{\pi^{3/2}}\lambda-\frac{2}{3\pi^{3/2}}\lambda^3+O(\lambda^5) \right)
\label{eq:d0mp2nd}
\end{equation}

\begin{table}
  \begin{center}

\def~{\hphantom{0}}

\begin{tabular}{ c | c }
$g(x)$ &  $h(x)$ \\
\hline
$1$ &  $\sqrt{1-x^2}$ \\
$\delta''_D(x)$ & $\displaystyle\frac{1}{\pi}\,\frac{\sqrt{1-x^2}}{x^2}$ \\[8 pt]
$\delta'_D(x$) & $-\displaystyle\frac{1}{\pi}\frac{\sqrt{1-x^2}}{x}$ \\[8 pt]
$\delta_D(x)$ & $\displaystyle\frac{1}{\pi}\,\text{atanh}\sqrt{1-x^2}$ \\[8 pt]
$\text{sign}(x)/2$ & $\displaystyle\frac{1}{\pi}x\,\text{atanh}\sqrt{1-x^2}$ \\[8 pt]
$|x|-\displaystyle\frac{1}{\pi}$ &  $\displaystyle\frac{1}{\pi}x^2\,\text{atanh}\sqrt{1-x^2}$ \\[8 pt]
$|x|x$ & $\displaystyle\frac{2}{3\pi}x^3\,\text{atanh}\sqrt{1-x^2}+\displaystyle\frac{2}{3\pi}x\sqrt{1-x^2}$\\[8 pt]
$|x|^3-\displaystyle\frac{1}{3\pi}$ &  $\displaystyle\frac{1}{2\pi}x^4\,\text{atanh}\sqrt{1-x^2}+\displaystyle\frac{1}{2\pi}x^2\sqrt{1-x^2}$\\[8 pt]
\end{tabular}

  \caption{Select solutions $h(x)$ to the problem $g(x)=\frac{1}{\pi}\int_{-1}^1\frac{h'(s)}{x-s}ds$, with $h(\pm1)=0$.}
  \label{tab:kd}
  \end{center}
\end{table}

\section{Outer solution in the critically stressed limit}
\label{sec:6}

\medskip

We now look for an asymptotic expansion of the solution in the critically stressed limit, in which the rupture front outpaces fluid pressure diffusion, $\lambda\gg 1$. As noted by GG12 for their problem, the solution consists of an outer solution on distances comparable to the rupture distance $a(t)$ and an inner solution on distances comparable to $\sqrt{\alpha t}$. To look for the outer solution we solve for the slip distribution satisfying eq. (\ref{eq:nond}) after expanding the two terms on the left hand side in the large $\lambda$ limit. We begin by considering the expansion of the following function as $\lambda\rightarrow \infty$
\begin{equation}\frac{\text{erfc}\,|u|}{\sqrt{1-(u/\lambda)^2}}\approx \text{erfc}\,|u|+\frac{1}{2\lambda^2}u^2\text{erfc}\,|u|+O(\lambda^{-4})\label{eq:erfsqrt}\end{equation}
This function appears in the integral (\ref{eq:Kint}) following the change of variable $u=\lambda \bar x$, such that
\begin{equation}
\left(1-\frac{\tau}{\tau_p}\right)\frac{\sigma'}{\Delta p}=\frac{1}{\pi\lambda}\int_{-\lambda}^{\lambda}\frac{\text{erfc}|u|}{\sqrt{1-(u/\lambda)^2}}du\approx\frac{2}{\pi^{3/2}}\frac{1}{\lambda} +\frac{1}{3\pi^{3/2}}\frac{1}{\lambda^3}+O(\lambda^{-5})
\label{eq:Kasy2}
\end{equation}
In addition, we perform a multipole expansion of the distribution (Appendix \ref{appC})
\begin{equation}
\text{erfc}\,|\lambda \bar x| \approx p_0\delta_D (\bar x)-p_1\delta_D'(\bar x)+p_2  \delta''_D(\bar x)+O(\lambda^{-5})
\label{eq:mpole}
\end{equation}
where $\delta_D$ is the Dirac delta and its first and second derivatives, $\delta_D'(x)$ and $\delta_D''(x)$, with the properties 
$$\int_{-\infty}^\infty \delta_D(x)dx=1,\,\qquad \int_{-\infty}^\infty x\delta'_D(x)dx=1,\qquad \int_{-\infty}^\infty \frac{x^2}{2}\delta''_D(x)dx=1$$
and where the coefficients
\begin{align}
p_0&=\int_{-1}^1 \text{erfc}\,|\lambda\bar x|\,d\bar x=\frac{1}{\lambda}\int_{-\lambda}^\lambda \text{erfc}\,|\hat x|\,d\hat x=\frac{2}{\sqrt{\pi}}\frac{1}{\lambda}+O(\exp(-\lambda^2)/\lambda)\\ 
p_1&=\int_{-1}^1 \bar x \text{erfc}\,|\lambda\bar x|\,d\bar x=0\\
p_2&=\int_{-1}^1 \frac{\bar x^2}{2} \text{erfc}\,|\lambda\bar x|\,d\bar x=\frac{1}{3\sqrt{\pi}}\frac{1}{\lambda^3}+O(\exp(-\lambda^2)/\lambda^{3})\label{eq:coeff2}
\end{align}
exhibit beyond-all-orders decay at large $\lambda$ at the rate $\exp(-\lambda^2)$, following the leading-order term.
With eqs. (\ref{eq:Kasy2}) and (\ref{eq:mpole}), the equation governing the slip distrbution (\ref{eq:nond}) becomes 
\begin{equation}
\frac{1}{\pi}\int_{-1}^{1}\frac{d\bar\delta/d\bar s}{\bar x-\bar s}d\bar s \approx \frac{1}{\lambda}\left(-\frac{2}{\pi^{3/2}} + \frac{2}{\sqrt{\pi}}\delta_D(\bar x) \right) +\frac{1}{\lambda^3} \left(-\frac{1}{3\pi^{3/2}} + \frac{1}{3\sqrt{\pi}}\delta''_D(\bar x) \right) +O(\lambda^{-5})
\end{equation}

\medskip

As for the marginally pressurized case, we look for a solution in the form of a perturbation expansion
\begin{equation}
\bar \delta(\bar x)\approx \frac{1}{\lambda}\delta_0(\bar x)+\frac{1}{\lambda^3}\delta_1(\bar x)+O(\lambda^{-5})
\label{eq:asyexp2}
\end{equation}
where $\delta_0$ and $\delta_1$ now satisfy
\[ \frac{1}{\pi}\int_{-1}^{1}\frac{d\delta_0/d\bar s}{\bar x-\bar s}d\bar s=
-\frac{2}{\pi^{3/2}}+\frac{2}{\sqrt{\pi}}\delta_D(\bar x),
\quad \frac{1}{\pi}\int_{-1}^{1}\frac{d\delta_1/d\bar s}{\bar x-\bar s}d\bar s=
-\frac{1}{3\pi^{3/2}}+\frac{1}{3\sqrt{\pi}}\delta''_D(\bar x) \]
and the solutions
\begin{equation}\delta_0(\bar x)=\frac{2}{\pi^{3/2}}\left(\text{atanh}\, \sqrt{1-\bar x^2}-\sqrt{1-\bar x^2}\right)\label{eq:d0cs}\end{equation}
\begin{equation}\delta_1(\bar x)=\frac{1}{3\pi^{3/2}}\left(\frac{\sqrt{1-\bar x^2}}{\bar x^2}-\sqrt{1-\bar x^2}\right)
\label{eq:d1cs}\end{equation}
are again found by superposing solutions provided in Table \ref{tab:kd}. 

\medskip

From the above, the outer solution for slip near $\bar x =0$ has the behavior 
\begin{equation}
\delta(\bar x)\approx \frac{\lambda\sqrt{\alpha t}f\Delta p}{\mu'}\left[\frac{1}{\lambda}\frac{2}{\pi^{3/2}}\left(\log\frac{2}{|\bar x|}-1+\frac{\bar x^2}{4}+O(\bar x^4)\right)+\frac{1}{\lambda^3}\frac{1}{3\pi^{3/2}}\left(\frac{1}{\bar x^2}-\frac{3}{2}+O(\bar x^2)\right)+O(\lambda^{-5})\right]
\label{eq:inout}
\end{equation}

\section{Inner solution in the critically stressed limit}
\label{sec:7}

\medskip

To examine the behavior of slip on the diffusive lengthscale $\sqrt{\alpha t}$, we perform a change of variable to the scale distance $\hat x = \lambda \bar x = x/\sqrt{\alpha t}$, such that eq. (\ref{eq:dinv}) becomes
\begin{equation}
\lambda\frac{d\bar \delta}{d\hat x}=-\frac{\sqrt{1-(\hat x/\lambda)^2}}{\pi}\int_{-\lambda}^\lambda\frac{\text{erfc}|\hat s|}{\sqrt{1-(\hat s/\lambda)^2}}\frac{1}{\hat x-\hat s}d\hat s
\label{eq:dinv1}
\end{equation}
Rescaling slip as $\hat \delta = \lambda \bar \delta = \delta /(\sqrt{\alpha t}f\Delta p/\mu')$, we perform a series expansion of the square-root terms for large $\lambda$
\begin{equation}
\frac{d\hat \delta}{d\hat x}\approx-\frac{1}{\pi}\left(1-\frac{1}{\lambda^2}\frac{\hat x^2}{2}+O(\lambda^{-4})\right)\int_{-\lambda}^\lambda\left(\text{erfc}\,|\hat s|+\frac{1}{\lambda^2}\frac{\hat s^2}{2}\text{erfc}\,|\hat s|+O(\lambda^{-4})\right)\frac{1}{\hat x-\hat s}d\hat s
\end{equation}
and regroup terms of similar order
\begin{equation}
\frac{d\hat \delta}{d\hat x}\approx-\frac{1}{\pi}\int_{-\lambda}^\lambda\frac{\text{erfc}\,|\hat s|}{\hat x - \hat s}d\hat s+\frac{1}{\lambda^2}\left(\frac{\hat x^2}{2}\frac{1}{\pi}\int_{-\lambda}^\lambda \frac{\text{erfc}\,|\hat s|}{\hat x-\hat s}d\hat s-\frac{1}{\pi}\int_{-\lambda}^\lambda \frac{(\hat s^2/2)\text{erfc}\,|\hat s|}{\hat x - \hat s}d\hat s\right)+O(\lambda^{-4})
\end{equation}
Given the beyond-all-orders decay, for modestly large values $\hat s$, of the term $\text{erfc}\,|\hat s|\approx \exp(-\hat s^2)/(\sqrt{\pi}\hat s)$ appearing in all of the integrands above, the limits of the integrals may pass  to $\infty$ without consequence for the above expansion in powers of $\lambda$. Hence,
\begin{equation}
\frac{d\hat \delta}{d\hat x}\approx-\frac{1}{\pi}\int_{-\infty}^\infty\frac{\text{erfc}\,|\hat s|}{\hat x - \hat s}d\hat s+\frac{1}{\lambda^2}\left(\frac{\hat x^2}{2}\frac{1}{\pi}\int_{-\infty}^\infty \frac{\text{erfc}\,|\hat s|}{\hat x-\hat s}d\hat s-\frac{1}{\pi}\int_{-\infty}^\infty \frac{(\hat s^2/2)\text{erfc}\,|\hat s|}{\hat x - \hat s}d\hat s\right)+O(\lambda^{-4})
\label{eq:dininfty}
\end{equation}
The above integrals are Hilbert transforms, defined as
\[
\mathcal{H}[f(s)]=\frac{1}{\pi}\int_{-\infty}^\infty \frac{f(s)}{x-s}ds
\]
which have the property 
\begin{align*}
\mathcal{H}[s^2f(s)]&=\mathcal{H}[(s^2-x^2+x^2)f(s)]\\
&=x^2\mathcal{H}[f(s)]+\mathcal{H}[(s-x)(s+x)f(s)]\\
&=x^2\mathcal{H}[f(s)]-\frac{1}{\pi}\int_{-\infty}^{\infty}sf(s)ds-\frac{x}{\pi}\int_{-\infty}^{\infty}f(s)ds
\end{align*}
that, when applied to the third integral in eq. (\ref{eq:dininfty})
\begin{equation}
\frac{1}{\pi}\int_{-\infty}^\infty \frac{(\hat s^2/2)\text{erfc}\,|\hat s|}{\hat x - \hat s}d\hat s=\frac{\hat x^2}{2}\frac{1}{\pi}\int_{-\infty}^\infty \frac{\text{erfc}\,|\hat s|}{\hat x-\hat s}d\hat s -\frac{\hat x}{\pi^{3/2}}
\end{equation}
leads to the reduction of eq. (\ref{eq:dininfty}) to 
\begin{equation}
\frac{d\hat \delta}{d\hat x}\approx-\frac{1}{\pi}\int_{-\infty}^\infty\frac{\text{erfc}\,|\hat s|}{\hat x - \hat s}d\hat s+\frac{1}{\lambda^2}\frac{\hat x}{\pi^{3/2}} +O(\lambda^{-4})
\end{equation}
Subsequently integrating from 0 to $\hat x$, we find the inner solution to within a yet-undetermined constant $\delta(0)$
\begin{equation}
\hat \delta(\hat x) \approx \hat \delta (0) - \underline{\int_0^{\hat x}\left[\frac{1}{\pi }\int_{-\infty}^\infty \frac{\text{erfc}\,|s|}{x -s}d s\right]dx} + \frac{1}{\lambda^2}\frac{\hat x^2}{2\pi^{3/2}}+O(\lambda^{-4})
\label{eq:din}
\end{equation}

\medskip

\begin{figure}
\begin{center}
\centerline{\includegraphics[scale=.75]{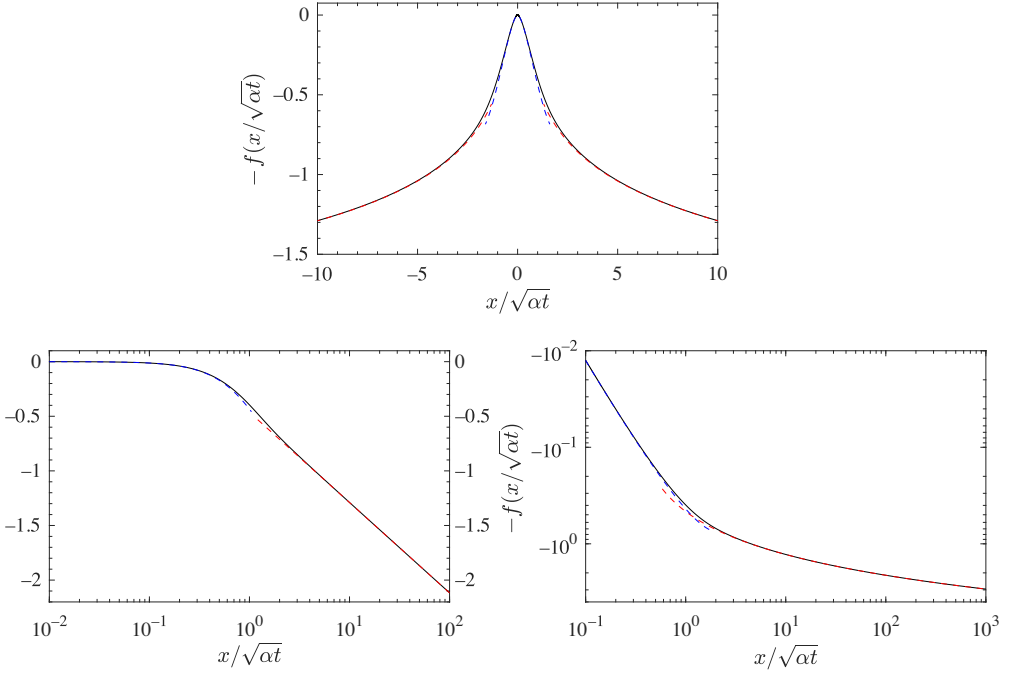}}
\end{center}
\caption{Spatio-temporal component, $f(x/\sqrt{\alpha t})$, of ``inner" solution for slip $\delta$ in the critically stressed limit, eq. (\ref{eq:in}). Shown in black on linear ({\bf top}), log-linear ({\bf left}), and logarithmic ({\bf right}) axes. Red- and blue-dashed curves: outer and inner asymptotic behavior of $f$, eqs (\ref{eq:fbg}) and (\ref{eq:fsm}). }
\label{fig6}
\end{figure}

We define the underlined term as the function $f(\hat x)$ whose first derivative is the Hilbert transform
\[ f'(\hat x) = \frac{1}{\pi}\int_{-\infty}^\infty\frac{\text{erfc}\,|s| }{\hat x -s}ds \]
Since the transform commutes with derivatives, $[\mathcal{H}(g)]'=\mathcal{H}(g')$, we find that $f''$  has the concise expression
\[ f''(\hat x)=-\frac{2}{\pi^{3/2}}\int_{-\infty}^\infty\frac{\text{sign}(s)\exp(-s^2) }{\hat x -s}ds \]
which may be further simplified
\begin{align}
f''(\hat x)&=-\frac{2}{\pi^{3/2}}\exp(-\hat x^2)\int_0^\infty \frac{\exp(\hat x^2-s^2)}{\hat x^2-s^2}2sds\notag\\
&=-\frac{2}{\pi^{3/2}}\exp(-\hat x^2)\int_{-\hat x^2}^\infty\frac{\exp(-u)}{u}du\notag \\
&=-\frac{2}{\pi^{3/2}}\exp(- \hat x^2)\text{Ei}(\tilde x^2) \label{eq:f''}
\end{align}
with the change of variable $u=s^2-\hat x^2$ used in the intermediate step. In Fig. \ref{fig6} we plot $f(\hat x)$ found by numerically integrating twice the expression for $f''(\hat x)$, eq. (\ref{eq:f''}).

\medskip

For $\hat x$ near 0, 
\[ f''(\hat x)=-\frac{2}{\pi^{3/2}}\left(\ln|\hat x|+\gamma\right)+O(\hat x^2) \]
and upon integrating twice with $f(0)=f'(0)=0$, the behavior of $f$ in this region is
\[ f(\hat x)=\frac{2}{\pi^{3/2}}\hat x^2\left(\ln\frac{1}{|\hat x|}-\frac{\gamma}{2}+\frac{3}{2}\right)+O(\hat x^4\ln|\hat x|) \]

For large $\hat x$ 
\[ f''(\hat x)=-\frac{2}{\pi^{3/2}}\left(\frac{1}{\hat x^2}+\frac{1}{\hat x^4}\right)+O(\hat x^{-6}) \]
Upon integrating twice with the condition that $f'(\infty)=0$,
\begin{equation}
f(\hat x)= c + \frac{2}{\pi^{3/2}}\left(\ln{|\hat x|}-\frac{1}{6\hat x^2}\right) +O(\hat x^{-4})
\label{eq:fxlarge}
\end{equation}
where $c$ is a yet-undetermined constant of integration.

\medskip

\begin{figure}
\centerline{\includegraphics[scale=.75]{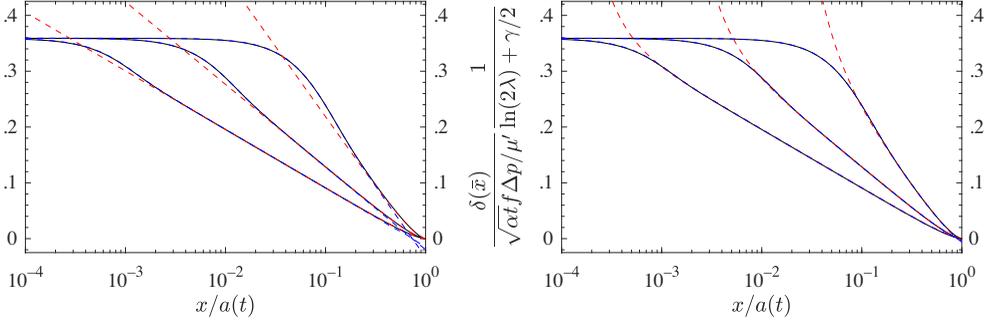}}
\caption{Self-similar distribution of slip in the critically stressed limit (large $\lambda$). In this limit, the rupture extent $a(t)$ outpaces the diffusive distance $\sqrt{\alpha t}$. ({\bf left and right}) Black curves from right to left correspond to full solutions self-similar slip profiles for $\lambda=10,100,1000$. ({\bf left}) Red-dashed and blue-dashed correspond to outer and inner solution expansion to first order. ({\bf right}) Red-dashed and blue-dashed correspond to outer and inner solution expansion to second order. Outer solution is given by eqs. (\ref{eq:asyexp2}--\ref{eq:d1cs}). Inner solution given by eq. (\ref{eq:din}).}
\label{fig7}
\end{figure}

We determine the constant $c$ following an approach used by \citet{Garagash12}, in which the order of integration in eq. (\ref{eq:din}) is swapped leading to an alternative expression for $f(\hat x)$
\begin{equation}
f(\hat x)=\int_0^{\hat x}\left[\frac{1}{\pi }\int_{-\infty}^\infty \frac{\text{erfc}\,|s|}{x -s}d s\right]dx=\frac{1}{\pi }\int_{-\infty}^\infty \text{erfc}\,|s| \,\ln\left |1-\frac{\hat x}{s}\right | ds
\label{eq:dinswap}
\end{equation}
and the expansion $\ln|1-\hat x/s |=\ln|\hat x| - \ln|s|+\ln|s/\hat x-1|$ is used to decompose the latter integral for $f(\hat x)$ into the sum
\begin{align}
f(\hat x) &= \frac{1}{\pi }\int_{-\infty}^\infty \text{erfc}\,|s|ds \ln|\hat x|  -\frac{1}{\pi }\int_{-\infty}^\infty \text{erfc}\,|s|\ln|s|ds+\frac{1}{\pi }\int_{-\infty}^\infty \text{erfc}\,|s|\ln\left | \frac{s}{\hat x}-1\right |ds\\
&=\frac{2}{\pi^{3/2}}\ln|\hat x| +\frac{2}{\pi^{3/2}}\left(1+\frac{\gamma}{2}\right)+\frac{1}{\pi }\int_{-\infty}^\infty \text{erfc}\,|s|\ln\left | \frac{s}{\hat x}-1\right |ds
\label{eq:falt}
\end{align}
Comparing (\ref{eq:fxlarge}) and (\ref{eq:falt}), we see that the asymptotic behavior of the last integral in (\ref{eq:falt}), for large $\hat x$, is given by the terms in (\ref{eq:fxlarge}), excluding the constant and the logarithmic terms. We also retrieve the value of the constant $c$
\[ c=\frac{2}{\pi^{3/2}}\left(1+\frac{\gamma}{2}\right) \]
\medskip
from which we conclude that the inner solution for slip has the asymptotic behavior at large $\hat x$
\begin{equation}
 \delta(\hat x) \approx \delta(0)-\frac{\sqrt{\alpha t}f\Delta p}{\mu'}\left[\frac{2}{\pi^{3/2}}\left(1+\frac{\gamma}{2}+\ln{|\hat x|}-\frac{1}{6|\hat x|^2}+O(\hat x^{-4})\right) - \frac{1}{\lambda^2}\frac{\hat x^2}{2\pi^{3/2}}  + O(\lambda^{-4})\right]
 \label{eq:outin}
\end{equation}

\medskip

\begin{figure}
\centerline{\includegraphics[scale=.75]{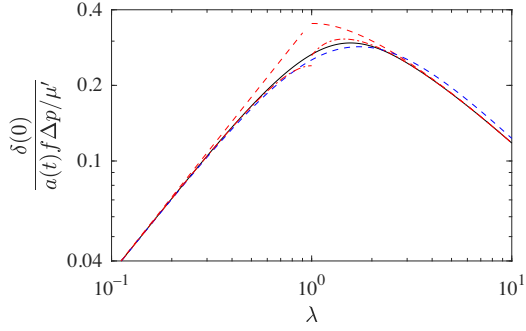}}
\caption{Comparison of approximations to the full solution of peak fault slip (black curve) in the vicinity of $\lambda=1$. First- and second- order approximations in the marginally pressurized and critically stressed limits, given respectively by eqs. (\ref{eq:d0mp2nd}) and (\ref{eq:d0crit2nd}), are shown as red-dashed and red-dot-dashed curves. Blue-dashed curve: approximation provided for all values of $\lambda$, eq. (\ref{eq:d0approx}).}
\label{fig8}
\end{figure}

\section{Matching inner and outer solutions}
\label{sec:8}

\medskip

We match the outer and inner solutions by equating eqs. (\ref{eq:inout}) and (\ref{eq:outin}) and solving for $\delta(0)$, the slip at the center in the critically stressed limit
\begin{equation}
\delta(0)\approx\frac{\sqrt{\alpha t}f\Delta p}{\mu'}\frac{2}{\pi^{3/2}}\left[\frac{\gamma}{2}+\ln(2\lambda)-\frac{1}{4\lambda^2}+O(\lambda^{-4})\right]
\label{eq:d0crit2nd}
\end{equation}
which completes the expression for the inner solution (\ref{eq:din}). In Fig. \ref{fig7}, we overlay the inner and outer solutions, to first and second order, above the full solutions for several large values of $\lambda$. The intermediate matching of the solutions is evident in the overlap of the dashed curves. As an aside, we can now show approximations to the slip at the center to first and second order in both the critically stressed and marginally pressurized limits in Fig. \ref{fig8}. For comparison, we also show the full solution, as well as the ad hoc approximation (\ref{eq:d0approx}) constructed using the first-order asymptotics.

\medskip

\begin{figure}
\centerline{\includegraphics[scale=.75]{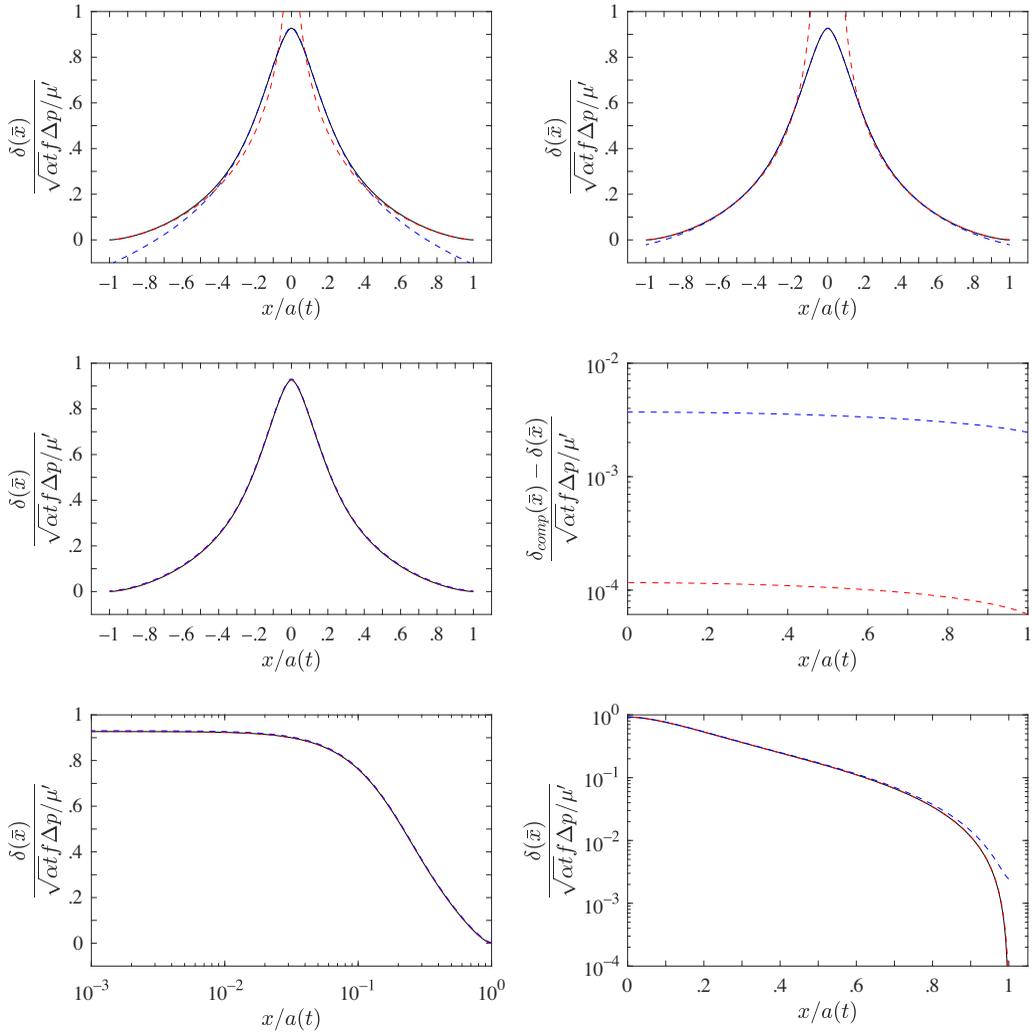}}
\caption{Solution for self-similar slip profile in black for $\lambda=5$. ({\bf top}) superimposed blue- and red-dashed curves, respectively, showing inner and outer solutions to ({\bf top-left}) leading and ({\bf top-order}) next order. ({\bf middle-left}) Superposition of leading and next-order composite solutions as blue- and red-dashed curves ({\bf middle-right}). The difference between the self-similar solution and the composite solutions at leading- and next-order, respectively blue and red-dashed curves. ({\bf bottom}) Semi-logarithmic plots comparing the leading- and next-order composite solutions, in blue- and red-dashed, to the self-similar solution, in black.}
\label{fig9}
\end{figure}

Using the inner and outer solutions, we construct a composite approximation \citep[e.g.,][]{Hinch91}
\[ \delta_{comp}(\bar x)=\delta_{in}(\lambda \bar x)+\delta_{out}(\bar x)-\delta_{overlap}(\bar x) \]
where $\delta_{in}$ and $\delta_{out}$ are the inner and outer solutions, eqs. (\ref{eq:din}) and (\ref{eq:asyexp2}), and
\[ \delta_{overlap}(\bar x)=\frac{2}{\pi^{3/2}}\left(\log\frac{2}{\bar x}-1+\frac{\bar x^2}{4}\right)+\frac{1}{\lambda^2} \frac{1}{3\pi^{3/2}}\left(\frac{1}{\bar x^2}-\frac{3}{2}\right)+O(\lambda^{-4}) \]
is their common intermediate form. In Fig. \ref{fig9} we compare the full numerical solution against the inner, outer, and composite solutions for a modestly large value of $\lambda=5$. The composite solution has an approximate error of O($\lambda^{-(n+2)}$) where $n=2\text{ or 4}$ is the order neglected in the asymptotic expansion.

\section{Summary and conclusion}

\medskip

The quasi-static rupture of a fault driven by a source of fluids has been studied in detail.
Tracking the rupture of the fault corresponds to a free boundary problem for which both the size of the slipping domain and the distribution of slip within must be solved. 
Both depend on a single dimensionless parameter whose limits correspond either to a fault whose initial, pre-injection shear stress is relatively close to the fault's pre-injection shear strength or to a fluid pressure increase that is marginally sufficient to induce sliding.
Because the problem involves contact between elastic half-spaces, interactions between points on the fault are non-local, in that slip in one location induces changes in the shear stress over the entire fault plane. The resulting equation governing slip evolution is an integro-differential equation.
In addition to the crack-tip boundary condition that slip vanish at the rupture front, the condition determining the free boundary is the absence of a stress singularity ahead of the rupture front, which corresponds to the boundary condition that the gradient of slip also vanishes at the rupture front.
Moreover, because the friction coefficient is held constant, there is an absence of an elasto-frictional lengthscale that may be otherwise present in problems for which friction depend explicitly on slip or its history.
Correspondingly, the only lengthscale in the problem is the diffusive length $\sqrt{\alpha t}$ with the consequence that spatial dependence of slip scales directly with this lengthscale, implying the self-similar propagation of slip.

\medskip

In a fashion similar to the earthquake-nucleation problem considered by GG12, we present asymptotic perturbation expansion solutions in the marginally pressurized and critically stressed limits and tabulate numerical solution for the intervening cases. In the critically stressed limit, the problem has both an inner solution on the scale of diffusion and an outer solution on the scale of the crack front. An advantage of the posed problem is that the marginally pressurized expansion and the outer solution of the critically stressed limit are expressible in concise closed form. For the latter, we develop a multipole expansion method to develop successive approximations of a distributed loading beyond a point-force approximation. The leading-order term of the inner solution of the critically stressed limit is solved numerically and tabulated. Key findings include that the rupture run-out distance from the point of injection follows $a(t)=\lambda\sqrt{\alpha t}$, where $\lambda$ is an amplification factor, originally solved for and presented by \citet{Bhattacharya19}. As noted there, if we denote the dimensionless problem parameter as $T\equiv (1-\tau/\tau_p)\sigma'/\Delta p$, then in the marginally pressurized limit, $\lambda \approx (\pi^{3/2}/4)(1-T)$ and in the critically stressed limit, $\lambda \approx (2/\pi^{3/2}) T^{-1}$. Furthermore, in the marginally pressurized limit, slip accumulates as $\delta\sim\lambda^2\sqrt{\alpha t} f\Delta p/\mu'$. In the critically stressed limit, slip accumulates as $\delta\sim\sqrt{\alpha t}f\Delta p/\mu'$ on inner distances of the order $\sqrt{\alpha t}$ from the injection point and as $\delta\sim\lambda^{-2}\sqrt{\alpha t}f\Delta p/\mu'$ at distances comparable to the rupture length $\lambda \sqrt{\alpha t}$.

\medskip

\backsection[Supplementary data]{\label{SupMat}Supplementary material is available at \\https://doi.org/10.1017/jfm.2019...}

\backsection[Funding]{This work was supported by the National Science Foundation (grant number 1653382).}

\backsection[Declaration of interests]{The authors report no conflict of interest.}

\backsection[Author ORCID]{R. C. Viesca, https://orcid.org/0000-0003-4180-7807}

\appendix

\section{Finite Hilbert transform solutions}\label{appA}
\medskip

The problem for slip posed as eq. (\ref{eq:nond}) has the form of a finite Hilbert transform 
\begin{equation}
g(x)=\frac{1}{\pi}\int_{-1}^1 \frac{h'(s)}{x-s}ds
\label{eq:fhilb0}
\end{equation}
where, here, $g(x)$ corresponds to a prescribed loading and $h$ of the distribution slip on the interface that is in quasi-static equilibrium with the loading.
The solution to eq. (\ref{eq:fhilb0}) is the known inversion \citep[e.g.,][]{Mushkhelishvili58, King09}
\begin{equation}
h'(x)= \frac{C}{\sqrt{1-x^2}}-\frac{1}{\sqrt{1-x^2}}\frac{1}{\pi}\int_{-1}^1\frac{\sqrt{1-s^2}g(s)}{x-s}ds\qquad C=\frac{1}{\pi}\int_{-1}^1h'(s)ds
\label{eq:fHilbinv}
\end{equation}
Since slip vanishes at the rupture boundaries, the corresponding condition on $h$ is $h(\pm1)=0$ and hence $C=0$.

\medskip

As an example solution, consider the distribution $g(x)=\delta _D(x)$, which is equivalent to a distribution of a point-force at the origin in the corresponding crack problem. The inversion for $h'(x)$ is
\begin{equation}
h'(x)=-\frac{1}{\sqrt{1-x^2}}\frac{1}{\pi}\int_{-1}^1\frac{\sqrt{1-s^2}\delta _D(s)}{x-s}ds=-\frac{1}{\pi}\frac{1}{x\sqrt{1-x^2}}
\end{equation}
and integrating again with respect to $x$, with the condition $h(\pm1)=0$, we find that
\begin{equation}
h(x)=\frac{1}{\pi}\text{arctanh}\sqrt{1-x^2}
\label{eq:exsol}
\end{equation}
In Table \ref{tab:kd}, we present a number of similarly derived solutions from which we draw in the main text.

\medskip 

From eq. (\ref{eq:fHilbinv}), we find that $h'(x)$ has the behavior in the limit $x\rightarrow\pm 1$ 
\begin{equation}
h'(x)=-\frac{1}{\sqrt{2(1\mp x)}}\frac{1}{\pi}\int_{-1}^1\sqrt{\frac{1\pm s}{1\mp s}}g(s)ds
\label{eq:htips}
\end{equation}
We may compare this behavior to the leading-order term in the \citet{Williams57} solution for slip near the tip of a crack located at $x=\pm a$
\begin{equation}
\delta(x)= \frac{K}{\mu'} \sqrt{\frac{2(a\mp x)}{\pi}}\qquad \frac{d\delta}{dx}=\mp \frac{K}{\mu'} \sqrt{\frac{1}{2(a\mp x)}}
\label{eq:SIF}
\end{equation}
where $K$ is the conventionally defined mode-II or mode-III stress intensity factor and the corresponding leading order term in the distribution of stress ahead of the tip is $\tau_{tip}(x)=K/\sqrt{2\pi(x\mp a)}$. We may define a quantity $k$ corresponding to the stress intensity factor $K$ by $k=K/(\mu'\sqrt{a})$, and in comparing the latter expression in (\ref{eq:SIF}) with eq. (\ref{eq:htips}), we can derive an analogous expression for an intensity factor $k_\pm$ at $x=\pm 1$
\[
k_{\pm}=\frac{1}{\sqrt{\pi}}\int_{-1}^1\sqrt{\frac{1\pm s}{1\mp s}}g(s)ds
\]
Requiring that this intensity factor vanish at both tips, hence implies that two conditions be satisfied by the distribution $g(s)$
\[ \int_{-1}^1\sqrt{\frac{1\pm s}{1\mp s}}g(s)ds =0 \]
which can be recast as the sum and the difference of these two conditions, leading respectively to
\begin{equation}
\int_{-1}^1\frac{g(s)}{\sqrt{1-s^2}}ds=0\qquad \int_{-1}^1\frac{sg(s)}{\sqrt{1-s^2}}ds=0
\label{eq:K0}
\end{equation}
In the problem for the slip distribution, eq. (\ref{eq:nond}), we identify
\begin{equation}
g(s)=\text{erfc}\,|\lambda s| - \left(1-\frac{\tau}{\tau_p}\right)\frac{\sigma'}{\Delta p}
\label{eq:gs}
\end{equation}
The first condition of (\ref{eq:K0}) to be satisfied by (\ref{eq:gs}) corresponds to eq. (\ref{eq:Kint}) in main text, which provides the direct relation between $\lambda$ and the problem parameter $(1-\tau/\tau_p)\sigma'/\Delta p$. The second condition of (\ref{eq:K0}) is trivially satisfied by (\ref{eq:gs}), since $g(s)$ is an even function for that case.

\medskip

The non-singular stress conditions (\ref{eq:K0}), if present, can be taken into account in the inversion for $h'(x)$, eq. (\ref{eq:fHilbinv}),  which may be rewritten as
\begin{align}
h'(x) &= -\frac{1}{\sqrt{1-x^2}}\frac{1}{\pi}\int_{-1}^1\frac{\sqrt{1-s^2}g(s)}{x-s}ds+\frac{1}{\sqrt{1-x^2}}\frac{1}{\pi}\int_{-1}^1\frac{(x+s)g(s)}{\sqrt{1-s^2}}ds\notag\\
 &= -\frac{\sqrt{1-x^2}}{\pi}\int_{-1}^1\frac{1}{\sqrt{1-s^2}}\frac{g(s)}{x-s}ds\label{eq:invnonsing}
 \end{align}
Upon substituting eq. (\ref{eq:gs}) in eq. (\ref{eq:invnonsing}), the contribution of the constant parameter $(1-\tau/\tau_p)\sigma'/\Delta p$ vanishes, and we retrieve eq. (\ref{eq:dinv}) in the main text.

\medskip

As another example, we again consider the distribution $g(x)=\delta_D(x)$, for which the inversion for $h'(x)$ with the non-singular condition on $h'(x)$ is 
\[ h'(x)=\frac{-\sqrt{1-x^2}}{\pi}\int_{-1}^1\frac{1}{\sqrt{1-s^2}}\frac{\delta _D(s)}{x-s}ds=-\frac{1}{\pi}\frac{\sqrt{1-x^2}}{x}=-\frac{1}{\pi}\left(\frac{1}{x\sqrt{1-x^2}}-\frac{x}{\sqrt{1-x^2}}\right) \]
\begin{equation}
h(x)=\frac{1}{\pi}\left(\text{arctanh}\sqrt{1-x^2}-\sqrt{1-x^2}\right)
\label{eq:nonsingdD}
\end{equation}
to which we may compare, eq. (\ref{eq:exsol}), the inversion for the same distribution $g(x)$ without the non-singular condition. 

\medskip

In writing an asymptotic expansion for slip in powers $\lambda$ in the marginally pressurized and critically stressed limits, we derived spatial distributions for slip at each order ($\delta_0$ and $\delta_1)$ using results presented in Table \ref{tab:kd}. Note that the compilation of inversions in Table \ref{tab:kd} used (\ref{eq:fHilbinv}) and did not incorporate the non-singular crack conditions (\ref{eq:K0}) in the inversion for $h'(x)$ from $g(x)$. However, when solving for $\delta_0$ and $\delta_1$, the non-singular conditions are implicitly incorporated in the expansion for the stress parameter $(1-\tau/\tau_p)\sigma'/\Delta p$ in terms of $\lambda$. The resulting expressions for $\delta_0$ and $\delta_1$ are fully equivalent to the solutions that would have been found had the conditions (\ref{eq:K0}) been incorporated directly into the inversion. In other words, we would have arrived to the same expressions for $\delta_0$ and $\delta_1$ in Sections \ref{sec:5} and \ref{sec:6} had we instead used the non-singular inversion (\ref{eq:nonsingdD}) to construct the functions $h(x)$ in Table \ref{tab:kd} and subsequently applied them to solve for $\delta_0$ and $\delta_1$. We recognize this, for instance, in noting that the first term in the expansion for slip in the critically-stressed limit, eq. (\ref{eq:d0cs}), is given, to within the factor $p_0=2/\sqrt{\pi}$ by eq. (\ref{eq:nonsingdD}).

\section{Numerical solution procedure}\label{appB}

\medskip

To numerically solve for the slip distribution $\bar \delta(\bar x)$ from eq. (\ref{eq:dinv}), given a value of $\lambda$, we use Gauss-Chebyshev quadrature for singular integrals \citep{Erdogan73,Viesca18}. We begin by separating the singular integral in eq. (\ref{eq:dinv})
\begin{equation}
\phi(\bar z)=-\frac{1}{\pi}\int_{-1}^1 \frac{\text{erfc}|\lambda \bar s|}{\sqrt{1-\bar s^2}}\frac{1}{\bar z - \bar s}ds
\label{eq:interm}
\end{equation}
and note that the slip distribution is given by the integration
\begin{equation}
\bar \delta(\bar x)=\int_{-1}^{\bar x} \sqrt{1-\bar z^2}\phi(\bar z)d\bar z
\label{eq:intinterm}
\end{equation}
We numerically approximate the integral (\ref{eq:interm}) at a discrete set of points $z_i$ 
\begin{equation}
\phi(\bar z_i)\approx-\frac{1}{n}\sum_{j=1}^n \frac{\text{erfc}|\lambda \bar s_j|}{\bar z_i - \bar s_j}
\end{equation}
where the quadrature points $\bar s_j$ and $\bar z_i$ are
\begin{alignat}{2}
\bar z_i&=\cos(\pi i/n)  &i=1,\,...\,,\,n-1\\ \bar s_j&=\cos[\pi(j-1/2)/n] &j=1,\,...\,,\,n\notag
\end{alignat}
We again use Gauss-Chebyshev quadrature \citep{Mason02} to approximate the integral (\ref{eq:intinterm}) at a set of points $x_k$
\begin{equation}
\bar \delta(\bar x_k)\approx \frac{1}{n}\sum_{i=1}^{n-1}(1-\bar z^2_i)\phi(\bar z_i)H(\bar x_k-\bar z_i)
\end{equation}
where $\bar x_k=\bar s_k$ for $k=1,\,...\,,n$ and $H(\cdot)$ is the Heaviside step function.

\medskip

We may similarly approximate the relation between $(1-\tau/\tau_p)\sigma'/\Delta p$ and $\lambda$ by first prescribing a value of $\lambda$ and then numerically evaluating the integral (\ref{eq:Kint}) by Gauss-Chebyshev quadrature 
\begin{equation}
\left(1-\frac{\tau}{\tau_p}\right)\frac{\sigma'}{\Delta p}\approx \frac{1}{n}\sum_{j=1}^n\text{erfc}|\lambda s_j|
\end{equation}
where $s_j$ is defined as before.

\section{Multipole expansion}
\label{appC}
\medskip

Here we derive the multipole expansion of eqs. (\ref{eq:mpole}--\ref{eq:coeff2}). This expansion was used to derive the outer solution in the critically stressed limit, for which the rupture extent $a(t)\gg\sqrt{\alpha t}$, such that the fluid pressure source appears localized about the origin. We begin by noting that the solution to the problem for $h(x)$
\begin{equation}
g(x)=\frac{1}{\pi}\int_{-1}^1 \frac{h'(s)}{x-s}ds
\label{eq:fHilb}
\end{equation}
with the boundary conditions $h(-1)=h(1)=0$, may be also written in terms of the Green's function $G(x,x')$ satisfying
\[ \delta_D(x-x')=\frac{1}{\pi}\int_{-1}^1\frac{G(s,x')}{x-s}ds \]
Using the inversion eq. (\ref{eq:fHilbinv}), the Green's function is
\begin{align}
G(x,x')&=-\frac{1}{\sqrt{1-x^2}}\frac{1}{\pi}\int_{-1}^1\frac{\sqrt{1-s^2}\delta_D(s-x')}{x-s}ds\label{eq:Gf1}\\
&=-\frac{1}{\pi}\sqrt{\frac{1-x'^2}{1-x^2}}\frac{1}{x-x'}\notag
\end{align}
and the solution to eq. (\ref{eq:fHilb}) can be written as
\begin{equation}
h'(x)=\int_{-1}^1G(x,x')g(x')dx'
\label{eq:GFsoln}
\end{equation}

We derive a multipole expansion for eq. (\ref{eq:fHilb}) by considering that a Taylor series expansion of the Green's function in eq. (\ref{eq:GFsoln}) about the origin $x'=0$ 
\[ h'(x)=\int_{-1}^1\left[G(x,0)+x'\left.\frac{\partial G}{\partial x'}\right|_{x'=0}+\frac{x'^2}{2}\left.\frac{\partial^2 G}{\partial x'^2}\right|_{x'=0} +...+\frac{x'^n}{n!}\left.\frac{\partial ^nG}{\partial x'^n}\right|_{x'=0}\right]g(x')dx' \]
reduces to
\begin{equation}
h'(x)=p_0G(x,0)+p_1\left.\frac{\partial G}{\partial x'}\right|_{x'=0}+p_2\left.\frac{\partial^2 G}{\partial x'^2}\right|_{x'=0}+...+p_n\left.\frac{\partial^n G}{\partial x'^n}\right|_{x'=0}
\label{eq:mpolex1}
\end{equation}
where the coefficients of this series are
\begin{align*} p_0&=\int_{-1}^1g(x')dx',\quad p_1=\int_{-1}^1x'g(x')dx', \\p_2&=\int_{-1}^1\frac{x'^2}{2}g(x')dx',\quad...,\quad p_n=\int_{-1}^1\frac{x'^n}{n!}g(x')dx' \end{align*}

The expansion of $g(x)$ implied by eq. (\ref{eq:mpolex1}) is found by first noting that, from eq. (\ref{eq:Gf1}),
\begin{equation}
\frac{\partial^n G}{\partial x'^n}=-\frac{1}{\sqrt{1-x^2}}\frac{1}{\pi}\int_{-1}^1\frac{\sqrt{1-s^2}[(-1)^n\delta^{(n)}_D(s-x')]}{x-s}ds
\label{eq:mpoleG}
\end{equation}
where $\delta_D^{(n)}$ is the $n$-th derivative of $\delta_D$ with respect to its argument. When evaluating eq. (\ref{eq:mpoleG}) at $x'=0$ and comparing with eq. (\ref{eq:fHilbinv}), we see that eq. (\ref{eq:mpoleG}) is the inverted solution for $h'(x)$ when $g(x)=(-1)^n\delta^{(n)}_D(x)$, hence substituting eq. (\ref{eq:mpolex1}) into eq. (\ref{eq:fHilb}) yields
\[ g(x)=p_0\delta_D(x)-p_1\delta'_D(x)+p_2\delta''_D(x)-...+(-1)^np_n\delta^{(n)}_D(x) \]
where the leading two terms are the source monopole and dipole approximations, respectively. 
For $n=0,1,2$, the first few functions of $x$ in eq. (\ref{eq:mpolex1}) are 
\[ G(x,0)=-\frac{1}{\pi}\frac{1}{x\sqrt{1-x^2}},\quad \left.\frac{\partial G}{\partial x'}\right|_{x'=0}=-\frac{1}{\pi}\frac{1}{x^2\sqrt{1-x^2}},\quad \left.\frac{\partial^2 G}{\partial x'^2}\right|_{x'=0}=\frac{x^2-2}{x^3\sqrt{1-x^2}} \]
Multiplying these functions by $(-1)^n$ and integrating with respect to $x$, we find the expressions in Table \ref{tab:kd} for $h(x)$ when $g(x)=\delta_D(x)$ or one of its first two derivatives.

\bibliographystyle{jfm}

\begin{thebibliography}{99}

\expandafter\ifx\csname natexlab\endcsname\relax
\def\natexlab#1{#1}\fi
\expandafter\ifx\csname selectlanguage\endcsname\relax
\def\selectlanguage#1{\relax}\fi

\bibitem[Adachi and Detournay (2008)]{Adachi08}
{\sc Adachi, J. I., and Detournay, E.} 2008 {Plane strain propagation of a hydraulic fracture in a permeable rock}, {\it Eng. Frac. Mech.}, {\bf 75}, pp. 4666--4694.

\bibitem[Ball and Neufeld (2018)]{Ball18}
{\sc Ball, T. V. and Neufeld, J. A.} 2018 {Static and dynamic fluid-driven fracturing of adhered elastica}, {\it Phys. Rev. Fluids}, {\bf 3}, 074101.

\bibitem[Barenblatt (1956)]{Barenblatt56}
{\sc Barenblatt, G. I.} 1956 {On the formation of horizontal cracks in hydraulic fracture of an oil-bearing stratum}, {\it Prikl. Mat. Mech.}, {\bf 20}, pp. 475--486.

\bibitem[Bhattacharya and Viesca (2019)]{Bhattacharya19}
{\sc Bhattacharya, P., and Viesca, R. C.} 2019 {Fluid-induced aseismic fault slip outpaces pore-fluid migration}, {\it Science}, {\bf 364}, pp. 464--468.

\bibitem[Brantut (2021)]{Brantut21}
{\sc Brantut, N.} 2021 {Dilatancy toughening of shear cracks and implications for slow rupture propagation}, {\it arXiv}, 2104.06475.

\bibitem[Bunger and Cruden (2011)]{Bunger11}
{\sc Bunger, A. P., and Cruden, A. R.} 2011 {Modeling the growth of laccoliths and large mafic sills: role of magma body forces}, {\it J. Geophys. Res.}, {\bf 116}, B02203.

\bibitem[Ciardo and Lecampion (2019)]{Ciardo19}
{\sc Ciardo, F., and Lecampion, B.} 2019 {Effect of dilatancy on the transition from aseismic to seismic slip due to fluid injection in a fault}, {\it J. Geophys. Res.}, {\bf 124}, pp. 3724--3743.

\bibitem[Desroches et al. (1994)]{Desroches94}
{\sc Desroches, J., Detournay, E., Lenoach, B., Papanastasiou, P., Pearson, J. R. A., M. Thiercelin, and Cheng, A.} 1994 {The crack tip region in hydraulic fracturing}, {\it Proc. Roy. Soc. Lond. A}, {\bf 447}, pp. 39--48.

\bibitem[Detournay (2016)]{Detournay16}
{\sc Detournay, E.} 2016 {Mechanics of hydraulic fractures}, {\it Annu. Rev. Fluid Mech.}, {\bf 48}, pp. 311--339.

\bibitem[Dublanchet (2019)]{Dublanchet19}
{\sc Dublanchet, P.} 2019 {Fluid driven shear cracks on a strengthening rate-and-state frictional fault}, {\it J. Mech. Phys. Solids}, {\bf 132}, 103672.

\bibitem [Erdogan et al. (1973)]{Erdogan73}
{\sc  Erdogan F., Gupta, G.D., and Cook, T.S.} 1973 {Numerical solution of singular integral equations. Ch. 7 from } {\it Methods of Analysis and Solutions of Crack Problems}, {(ed. Sih, G.C.)}, {Mechanics of Fracture, vol. 1}, pp. 368--425, {Springer}.

\bibitem[Flitton and King (2004)]{Flitton04}
{\sc Flitton, J. C., and King, J. R.} 2004 {Moving boundary and fixed-domain problems for a sixth-order thin-film equation}, {\it Euro. J. Appl. Math.}, {\bf 15}, pp. 713--754.

\bibitem[Garagash (2012)]{Garagash12a}
{\sc Garagash, D. I.} 2012 {Seismic and aseismic slip pulses driven by thermal pressurization of pore fluid}, {\it J. Geophys. Res.}, {\bf 117}, B04314.

\bibitem[Garagash (2021)]{Garagash21}
{\sc Garagash, D. I.} 2021 {Fracture mechanics of rate-and-state faults and fluid injection induced slip}, {\it Phil. Trans. Roy. Soc. A}, {\bf 379}, 20200129.

\bibitem[Garagash and Detournay (2000)]{Garagash00}
{\sc Garagash, D. I., and Detournay, E.} 2000 {The tip region of a fluid-driven fracture in an elastic medium}, {\it J. Appl. Mech.}, {\bf 67}, pp. 183--192.

\bibitem[Garagash and Germanovich (2012)]{Garagash12}
{\sc Garagash, D. I., and Germanovich, L. N.} 2012 {Nucleation and arrest of dynamic slip on a pressurized fault}, {\it J. Geophys. Res.}, {\bf 117}, B10310.

\bibitem[Garipov et al. (2016)]{Garipov16}
{\sc Garipov, T. T., Karimi-Fard, M., and Tchelepi, H. A.} 2016 {Discrete fracture model for coupled flow and geomechanics}, {\it Comput. Geosci.}, {\bf 20}, pp. 149--160.

\bibitem[Healy et al. (1968)]{Healy1968}
{\sc Healy, J. H., Rubey, W. W., Griggs, D. R., and Raleigh, C. B.} 1966 {The Denver earthquakes}, {\it Science}, {\bf 161}, pp. 1301--1310.

\bibitem[Hewitt et al. (2015)]{Hewitt15}
{\sc Hewitt, I. J., Balmforth, N. J., and de Bruyn, J. R.} 2015 {Elastic-plated gravity currents}, {\it Euro. J. Appl. Math.}, {\bf 26}, pp. 1--31.

\bibitem[Hinch (1991)]{Hinch91}
{\sc Hinch, E. J.} 1991 {\it Perturbation Methods}, {Cambridge University Press}.

\bibitem[Hosoi and Mahadevan (2004)]{Hosoi04}
{\sc Hosoi, A. E., and Mahadevan, L.} 2004 {Peeling, healing, and bursting in a lubricated elastic sheet}, {\it Phys. Rev. Lett.}, {\bf 93}, 137802.

\bibitem[Jha and Juanes (2014)]{Jha14}
{\sc Jha, B., and Juanes, R.} 2014 {Coupled multiphase flow and poromechanics: A computational model of pore pressure effects on fault slip and earthquake triggering}, {\it Wat. Resour. Res.}, {\bf 50}, pp. 3776--3808.

\bibitem[Khristianovic and Zheltov (1955)]{Khristianovic55}
{\sc Khristianovic, S. A., and Zheltov, Y. P.} 1955 {Formation of vertical fractures by means of highly viscous fluid}, {\it Proc. 4th World Petrol. Congr.}, {\bf 1}, pp. 579--586, {World Petroleum Council}. 

\bibitem[Kennedy et al. (1997)]{Kennedy97}
{\sc Kennedy, B. M., Kharaka, Y. K., Evans, W. C., Ellwood, A., DePaolo, D. J., Thordsen, J., Ambats, G., and Mariner, R. H.} 1997 {Mantle fluids in the San Andreas fault system, California}, {\it Science.}, {\bf 278}, pp. 1278--1281.

\bibitem[King (2009)]{King09}
{\sc King, F. W.} 2009 {\it Hilbert Transforms}, vols. 1 \& 2, {Cambridge University Press}.

\bibitem[Lenoach (1995)]{Lenoach95}
{\sc Lenoach, B.} 1995 {The crack tip solution for hydraulic fracturing in a permeable solid}, {\it J. Mech. Phys. Solids}, {\bf 43}, pp. 1025--1043.

\bibitem[Lipovsky and Dunham (2017)]{Lipovsky17}
{\sc Lipovsky, B. P., and Dunham, E. R.} 2017 {Slow-slip events on the Whillans Ice PLain, Antarctica, described using rate-and-state friction as an ice stream sliding law}, {\it J. Geophys. Res.}, {\bf 122}, pp. 973--1003.

\bibitem[Lister (1990a)]{Lister90a}
{\sc Lister, J. R.} 1990 {Buoyancy-driven fluid fracture: the effects of material toughness and of low-viscosity precursors}, {\it J. Fluid Mech.}, {\bf 210}, pp. 263--280.

\bibitem[Lister (1990b)]{Lister90b}
{\sc Lister, J. R.} 1990 {Buoyancy-driven fluid fracture: similarity solutions for the horizontal and vertical propagation of fluid-filled cracks}, {\it J. Fluid Mech.}, {\bf 217}, pp. 213--239.

\bibitem[Lister et al. (2013)]{Lister13}
{\sc Lister, J. R., Peng, G. G., and Neufeld, J. A.} 2013 {Viscous control of peeling an elastic sheet by bending and pulling}, {\it Phys. Rev. Lett.}, {\bf 111}, 154501.

\bibitem[Lister et al. (2019)]{Lister19}
{\sc Lister, J. R., Skinner, D. J., and Large, T. M. J.} 2019 {Viscous control of shallow elastic fracture: peeling without precursors}, {\it J. Fluid. Mech.}, {\bf 868}, pp. 119-140.

\bibitem[Marck et al. (2015)]{Marck15}
{\sc Marck, J., Savitski, A. A., and Detournay, E.} 2015 {Line source in a poroelastic layer bounded by an elastic space}, {\it Int. J. Numer. Anal. Meth. Geomech.}, {\bf 39}, pp. 1484--1505.

\bibitem[Mason and Handscomb (2002)]{Mason02}
{\sc Mason, J. C., and Handscomb, D. C.} 2002 {\it Chebyshev Polynomials}, {Chapman and Hall}.

\bibitem[McClung (1979)]{McClung79}
{\sc McClung, D. M.} 1979 {Shear fracture precipitated by strain softening as a mechanism }, {\it J. Geophys. Res.}, {\bf 84}, pp. 3519--3526.

\bibitem[Michaut (2011)]{Michaut11}
{\sc Michaut, C. M.} 2011 {Dynamics of magmatic intrusions in the upper crust: theory and applications to laccoliths on Earth and the Moon}, {\it J. Geophys. Res.}, {\bf 116}, B05205.

\bibitem[Mushkhelishvili (1958)]{Mushkhelishvili58}
{\sc Mushkhelishvili, N. I.} 1958 {\it Singular Integral Equations}, {Springer}.

\bibitem[Noda et al. (2009)]{Noda09}
{\sc Noda, H., Dunham, E. M., and Rice, J. R.} 2009 {Earthquake ruptures with thermal weakening and the operation of major faults at low overall stress levels}, {\it J. Geophys. Res.}, {\bf 114}, pp. B07302.

\bibitem[Palmer and Rice (1973)]{Palmer73}
{\sc Palmer, A. C., and Rice, J. R.} 1973 {The growth of slip surfaces in the progressive failure of over-consolidated clay}, {\it Proc. R. Soc. Lon. A}, {\bf 332}, pp. 527--548.

\bibitem[Peacock (2001)]{Peacock01}
{\sc Peacock, S. M.} 2001 {Are the lower planes of double seismic zones caused by serpentine dehydration in subducting oceanic mantle?}, {\it Geology}, {\bf 29}, pp. 299--302.

\bibitem[Peng and Lister (2020)]{Peng20}
{\sc Peng, G. G., and Lister, J. R.} 2020 {Viscous flow under an elastic sheet}, {\it J. Fluid Mech.}, {\bf 905}, A30.

\bibitem[Platt et al. (2015)]{Platt15}
{\sc Platt, J. D., Viesca, R. C., and Garagash, D. I.} 2015 {Steadily propagating slip pulses driven by thermal decomposition}, {\it J. Geophys. Res.}, {\bf 120}, pp. 6558--6591.

\bibitem[Puzrin and Germanovich (2005)]{Puzrin05}
{\sc Puzrin, A. M., and Germanovich, L. N.} 2005 {The growth of shear bands in the catastrophic failure of soils}, {\it Proc. R. Soc. A}, {\bf 461}, pp. 1199--1228.

\bibitem [Rice (1968)]{Rice68}
{\sc Rice J. R.} 1968 {Mathematical analysis in the mechanics of fracture. Ch. 3 from } {\it Fracture}, {vol. 2}, {(ed. Liebowitz, H.)}, pp. 191--311, {Academic Press}.

\bibitem [Rice (1973)]{Rice73}
{\sc Rice J. R.} 1973 {The initiation and growth of shear bands } {\it Proceedings of the Symposium on the Role of Plasticity in Soil Mechanics}, {vol. 2}, {(ed. Palmer, A. C.)}, pp. 263--274, {Cambridge University Department of Engineering}.

\bibitem[Rice (2006)]{Rice06}
{\sc Rice, J. R.} 2006 {Heating and weakening of faults during earthquake slip}, {\it J. Geophys. Res.}, {\bf 111}, B05311.

\bibitem[Rubin (1995)]{Rubin95}
{\sc Rubin, A. M.} 1995 {Propagation of magma-filled cracks}, {\it Annu. Rev. Earth Planet. Sci.}, {\bf 23}, pp. 287--336.

\bibitem[Rutqvist et al. (2007)]{Rutqvist07}
{\sc Rutqvist, J., Birkholzer, J., Cappa, F., and Tsang, C.-F.} 2007 {Estimating maximum sustainable injection pressure during geological sequestration of CO2 using coupled fluid flow and geomechanical fault-slip analysis}, {\it Energ. Convers. Manage.}, {\bf 48}, pp. 1798--1807.

\bibitem[Schmitt et al. (2011)]{Schmitt11}
{\sc Schmitt, S. V., Segall, P., and Matsuzawa, T. } 2011 {Shear heating-induced thermal pressurization during earthquake nucleation}, {\it J. Geophys. Res.}, {\bf 116}, B06308.

\bibitem[Spence and Sharp (1985)]{Spence85}
{\sc Spence, D. A., and Sharp, P.} 1985 {Self-similar solutions for elastohydrodynamic cavity flow}, {\it Proc. R. Soc. Lond. A}, {\bf 400}, pp. 289--313.

\bibitem[Tsai and Rice (2010)]{Tsai10}
{\sc Tsai, V. C., and Rice, J. R.} 2010 {A model for turbulent hydraulic fracture and application to crack propagation at glacier beds}, {\it J. Geophys. Res.}, {\bf 114}, F03007.

\bibitem[Wang and Detournay (2018)]{Wang18}
{\sc Wang, Z.-Q. and Detournay, E.} 2018 {The tip region of a near-surface hydraulic fracture}, {\it J. Appl. Mech.}, {\bf 85}, 041010.

\bibitem[Williams (1957)]{Williams57}
{\sc Williams, M. L.} 1957 {On the stress distribution at the base of a stationary crack}, {\it J. Appl. Mech.}, {\bf 79}, pp. 109--114.

\bibitem[Viesca and Garagash (2015)]{Viesca15}
{\sc Viesca, R. C., and Garagash, D. I.} 2015 {Ubiquitous weakening of faults due to thermal pressurization}, {\it Nat. Geosci.}, {\bf 8}, pp. 875--879.

\bibitem[Viesca and Garagash (2018)]{Viesca18}
{\sc Viesca, R. C., and Garagash, D. I.} 2018 {Numerical methods for coupled fracture problems}, {\it J. Mech. Phys. Solids}, {\bf 113}, pp. 13--34.

\bibitem[Viesca and Rice (2012)]{Viesca12}
{\sc Viesca, R. C., and Rice, J. R.} 2012 {Nucleation of slip-weakeing rupture instability in landslides by localized increase of pore pressure}, {\it J. Geophys. Res.}, {\bf 117}, B03104.

\bibitem[Yang and Dunham (2021)]{Yang21}
{\sc Yang, Y. and Dunham, E. M.} 2021 {Effect of porosity and permeability evolution on injection-induced aseismic slip}, {\it J. Geophys. Res.}, {\bf 126}, e2020JB021258.

\bibitem[Zhu et al. (2020)]{Zhu20}
{\sc Zhu, W., Allison, K. L., Dunham, E. M., and Yang, Y.} 2020 {Fault valving and pore pressure evolution in simulations of earthquake sequences and aseismic slip}, {\it Nat. Comm.}, {\bf 11}, 4833.
\end{thebibliography}


\end{document}